\documentclass[11pt,a4paper]{PoS}

\usepackage{xspace,amsmath,slashed,mathrsfs,url,blindtext}

\newcommand{\pippi}{\textsf{pippi}\xspace}
\newcommand{\lagr}{\mathscr{L}}
\newcommand{\tm}{\textnormal}
\newcommand{\ovr}{\overline}

\title{Global study of effective Higgs portal dark matter models using GAMBIT}

\ShortTitle{Global study of effective Higgs portal dark matter models using GAMBIT}

\author{\speaker{Ankit Beniwal} \rm{(On behalf of the GAMBIT Collaboration)}\\
        Center for Cosmology, Particle Physics and Phenomenology (CP3), \\
		Universit\'{e} catholique de Louvain, B-1348 Louvain-la-Neuve, Belgium\\
        E-mail: \email{ankit.beniwal@uclouvain.be}, ORCID ID: \href{https://orcid.org/0000-0003-4849-0611}{\sf{\textrm{0000-0003-4849-0611}}}}

\abstract{In this proceeding, we present frequentist and Bayesian results from a global fit of effective vector and Majorana fermion Higgs portal dark matter (DM) models using the \textsf{GAMBIT} software. We systematically explore the parameter space of these models using advanced sampling techniques to simultaneously satisfy the observed DM abundance, Higgs invisible decay, and indirect and direct detection limits. In addition, we take account of a set of nuisance parameters arising from Standard Model, nuclear physics, DM halo and velocity distribution.~For the vector DM model viable solutions are found at low and high vector masses.~The Majorana fermion model requires a strong preference for a CP-odd, parity-violating coupling which leads to a momentum-suppression of the DM-nucleon cross-section.~All of our results, samples and input files are publicly available via \href{https://www.zenodo.org/communities/gambit-official/}{\textsf{Zenodo}}.}

\FullConference{
European Physical Society Conference on High Energy Physics - EPS-HEP2019 -\\
			10-17 July, 2019\\
			Ghent, Belgium}

\begin{document}

\section{Introduction}
The particle nature of dark matter (DM) remains a mystery despite a wealth of astrophysical/cosmological evidence to support its existence. A lack of particle DM candidates in the Standard Model (SM) forces us to look beyond the SM (BSM) \cite{Bertone:2010zza}. Multiple BSM extensions have been proposed, but Weakly Interacting Massive Particles (WIMPs) remain a popular choice. They appear in many BSM theories, and with a weak-scale interaction cross-section, they can also saturate the observed DM abundance.

In a bottom-up effective field theory (EFT) approach, one often tries to connect the DM and SM sectors via a portal-type interaction. In this regard, \emph{Higgs portal} models \cite{SilveiraZee,McDonald94,Burgess01,Davoudiasl2005,Patt:2006fw} are one of the most popular portals. They connect DM and SM particles via an $H^\dagger H$ operator. Not only is it one of the lowest-dimensional, gauge invariant operators in the SM, the discovery of a SM-like Higgs boson in 2012 at the LHC \cite{Aad:2012tfa,Chatrchyan:2012xdj} opens up a new window to directly probe its interaction with DM particles.

Here we present some frequentist and Bayesian results from a global study of the vector and Majorana fermion Higgs portal models \cite{Athron:2018hpc} using the \textsf{GAMBIT} \cite{Athron:2017ard} software. Using state-of-the-art sampling algorithms, we perform numerical scans of the model parameter space and find regions that simultaneously satisfy the observed DM relic density, Higgs invisible decay, and indirect and direct search limits.~We also systematically include a set of nuisance parameters from the SM, nuclear physics, and DM halo and velocity distribution. For more details, see Ref.~\cite{Athron:2018hpc}.

\section{Model details}
Assuming that the DM fields are SM gauge singlets, we consider vector $(V_\mu)$ and Majorana fermion $(\chi)$ DM particles.\footnote{The results for the Dirac fermion model are analogous to the Majorana fermion ones, thus we only present results for the Majorana fermion model; for more details, see Ref.~\cite{Athron:2018hpc}.}~A global $\mathbb{Z}_2$ symmetry: $(V_\mu, \chi) \rightarrow -(V_\mu, \chi)$ ensures the stability of our DM candidates on cosmological time scales. 

Our model Lagrangians after electroweak symmetry breaking and chiral rotation, $\chi \rightarrow e^{i\gamma_5 \alpha/2} \chi$, are \cite{Athron:2018hpc}
\begin{align}
	\lagr_V &= \lagr_{\tm{SM}} - \frac{1}{4} W_{\mu\nu} W^{\mu\nu} + \frac{1}{2} m_V^2 V_\mu V^\mu - \frac{1}{4!} \lambda_V (V_\mu V^\mu)^2 + \frac{1}{2} \lambda_{hV} V_\mu V^\mu \left(v_0 h + \frac{1}{2} h^2 \right), \\ 
	\lagr_\chi &= \lagr_{\tm{SM}} + \frac{1}{2} \ovr{\chi} (i \slashed{\partial} - m_\chi) \chi - \frac{1}{2} \frac{\lambda_{h\chi}}{\Lambda_\chi} (\cos\xi \, \ovr{\chi} \chi + \sin\xi \, \ovr{\chi} i \gamma_5 \chi) \left(v_0 h + \frac{1}{2} h^2 \right),
\end{align}
where $\lagr_{\tm{SM}}$ is the SM Lagrangian, $W_{\mu\nu} \equiv \partial_\mu V_\nu - \partial_\nu V_\mu$ is the vector field strength tensor, and $\lambda_{hV}$ $(\lambda_{h\chi}/\Lambda_\chi)$ is the dimensionless (dimensionful) coupling between the vector (Majorana fermion) DM and Higgs field.~The case $\xi = 0\,(\pi/2)$ corresponds to CP-even, parity-conserving (CP-odd, parity-violating) couplings. The physical masses of our DM candidates are:
\begin{equation}
	m_V = \sqrt{\mu_V^2 + \frac{1}{2} \lambda_{hV} v_0^2}\,, \quad
	m_\chi = \sqrt{ \left(\mu_\chi + \frac{1}{2} \frac{\lambda_{h\chi}}{\Lambda_\chi} v_0^2 \cos\theta \right)^2 + \left(\frac{1}{2} \frac{\lambda_{h\chi}}{\Lambda_\chi} v_0^2 \sin\theta \right)^2 }.
\end{equation}

\section{Constraints and likelihoods}\label{sec:theory}
The free parameters of the vector and Majorana fermion DM model, namely $\{m_V,\,\lambda_{hV}\}$ and $\{m_\chi,\,\lambda_{h\chi}/\Lambda_\chi,\,\xi\}$ respectively, are subject to a set of theoretical and observational constraints.\footnote{The quartic coupling $\lambda_V$ does not play any role in the DM phenomenology and can be ignored.} These are outlined below; for more details, see Ref.~\cite{Athron:2018hpc}.

\begin{enumerate}

	\item \textbf{Theoretical constraints}:~We require the perturbative unitarity of $VV \rightarrow hh$ scattering amplitudes, which translates into the following upper limit: $0 \leq \lambda_{hV} \leq 2 m_V^2/v_0^2$ \cite{Lebedev:2011iq}.
	
	The EFT validity of the Majorana fermion model depends on a specific UV completion.~By considering a simple UV completion \cite{Kim:2008pp}, we arrive at the following upper limit: $\lambda_{h\chi}/\Lambda_\chi < 4\pi/(2 m_\chi)$.\footnote{Above this value, the fermion DM annihilation cross-section $\sigma v_{\textnormal{rel}}$ may receive substantial corrections from new physics. This is not captured by our EFTs, and thus the resulting likelihoods/results should be interpreted with caution.}

	\item \textbf{Thermal relic abundance}:~We compute the DM relic density by numerically solving the Boltzmann equation at each parameter point.~In practice, we use the routines implemented in \textsf{DarkSUSY} \cite{Bringmann:2018lay} via \textsf{DarkBit} \cite{Workgroup:2017lvb}; the relevant annihilation cross-sections are presented in Appendix B of Ref.~\cite{Athron:2018hpc}.

	We \emph{do not} require our models to saturate the observed DM abundance. Thus, we use a one-sided Gaussian likelihood function for the DM relic density that is centered at the \emph{Planck} measured value: $\Omega_{\textnormal{DM}} h^2 = 0.1188 \pm 0.0010$ \cite{Ade:2015xua}. In doing so, we include a 5\% theoretical error on the computed values of the relic density, and combine it in quadrature with the \emph{Planck} measured uncertainty.

	\item \textbf{Higgs invisible decays}:~For $m_{V,\,\chi} < m_h/2$, the SM-like Higgs boson is kinematically allowed to decay into DM particles.~This contributes to the invisible decay width $\Gamma_{\textnormal{inv}}^h$; for exact analytical expressions, see Ref.~\cite{Athron:2018hpc}.

	For SM-like Higgs couplings, $\Gamma_{\textnormal{inv}}^h \leq 0.19~\Gamma^h_{\textnormal{total}}$ at $2\sigma$ C.L.~\cite{Belanger:2013xza}. This is taken into account in our study via \textsf{DecayBit} \cite{Workgroup:2017bkh}. Apart from the above limit, the LHC provides only a mild constraint on Higgs portal models \cite{Han:2016gyy}.

	\item \textbf{Indirect detection via gamma rays}:~We use a combined analysis of 15 dwarf galaxies using 6 years of \emph{Fermi}-LAT data with \texttt{Pass-8} event-level analysis \cite{Ackermann:2015zua}.~In practice, this uses a binned-likelihood as implemented in \textsf{DarkBit} via the \textsf{gamLike}\footnote{\url{https://gamlike.hepforge.org}} package.

	For the Majorana fermion model with a CP-conserving $(\xi = 0)$ interaction, the annihilation cross-section vanishes as the DM relative velocity $v$ approaches zero.~Thus, scenarios with $\xi \neq 0$ lead to non-trivial indirect detection limits.

	\item \textbf{Direct detection}:~For the vector DM model, the DM-nucleon cross-section is nuclear spin-independent (SI); for an analytical expression, see Ref.~\cite{Athron:2018hpc}.~On the other hand, the $\ovr{\chi} i\gamma_5 \chi$ term in the Majorana fermion model leads to a momentum-suppressed differential scattering cross-section \cite{Dienes:2013xya}, namely
	\begin{equation}\label{eqn:suppXsection}
		\frac{d\sigma_{\textnormal{SI}}^\chi}{dq^2} = \frac{1}{v^2} \left(\frac{\lambda_{h\chi}}{\Lambda_\chi}\right)^2 \frac{A^2 F^2 (E) \, f_N^2 \, m_N^2}{4\pi \, m_h^4} \left[ \cos^2 \xi + \frac{q^2}{4 m_\chi^2} \sin^2 \xi \right].
	\end{equation}
	Here $A$ is the mass number of the target isotope, $F^2(E)$ is the standard SI form factor, $m_N$ is the nucleon mass and $f_N$ is the effective Higgs-nucleon coupling.~As $|q| \simeq$ (1 -- 100) MeV $\ll m_\chi$, direct detection limits are significantly suppressed when $\xi \simeq \pi/2$.
	
	We use the \textsf{DarkBit} interface to \textsf{DDCalc 2.0.0}\footnote{\url{https://ddcalc.hepforge.org}} to compute a Poisson-based likelihood for the latest XENON1T 2018, LUX 2016, PandaX 2016 and 2017, CDMSlite, CRESST-II, PICO-60 and DarkSide-50 experiments.~The detector efficiencies and acceptance rates are modelled using pre-computed tables in \textsf{DDCalc}.~For more details, see Appendix A of Ref.~\cite{Athron:2018hpc}.

	\item \textbf{DM capture and annihilation in the Sun}:~We also include likelihoods from the 79-string IceCube searches for high-energy neutrinos from DM annihilation in the Sun \cite{Aartsen:2012kia} using \textsf{nulike} \cite{Aartsen:2016exj} via \textsf{DarkBit}. This uses a full unbinned likelihood based on the event-level energy and angular information of candidate events. The flavour and energy distributions of neutrinos are obtained using \textsf{WimpSim} \cite{Blennow:2007tw}, whereas the capture rates are computed using \textsf{Capt'n General}.\footnote{\url{https://github.com/aaronvincent/captngen}}

\end{enumerate}

All of the above constraints depend on a set of \emph{nuisance parameters}, i.e., parameters not appearing in our model Lagrangians but are required for the computation of various DM observables and likelihoods. Examples include the hadronic matrix elements, DM halo and velocity distribution parameters, and SM masses/couplings.~In our global fits, these are constrained by new likelihood terms that characterise their uncertainty. For a full list of nuisance parameters that are included in our study, see Ref.~\cite{Athron:2018hpc}.

\section{Scan details and results}
We perform parameter scans of the vector and Majorana fermion DM models.~For sampling from the posterior distribution, we use \textsf{T-Walk}, an ensemble Markov Chain Monte Carlo (MCMC) algorithm.~To find and explore the best-fit regions of our multi-dimensional likelihood function, we use \textsf{Diver}, a differential evolution sampler. For more details on these scanners, see Ref.~\cite{Workgroup:2017htr}.

In the following two subsections, we present profile likelihood and marginalised posterior plots in the planes of vector and Majorana fermion DM models, respectively. These are generated using \pippi \cite{Scott:2012qh}.

\subsection{Vector dark matter}
\begin{figure}[t]
	\centering
	
	\includegraphics[width=0.48\textwidth]{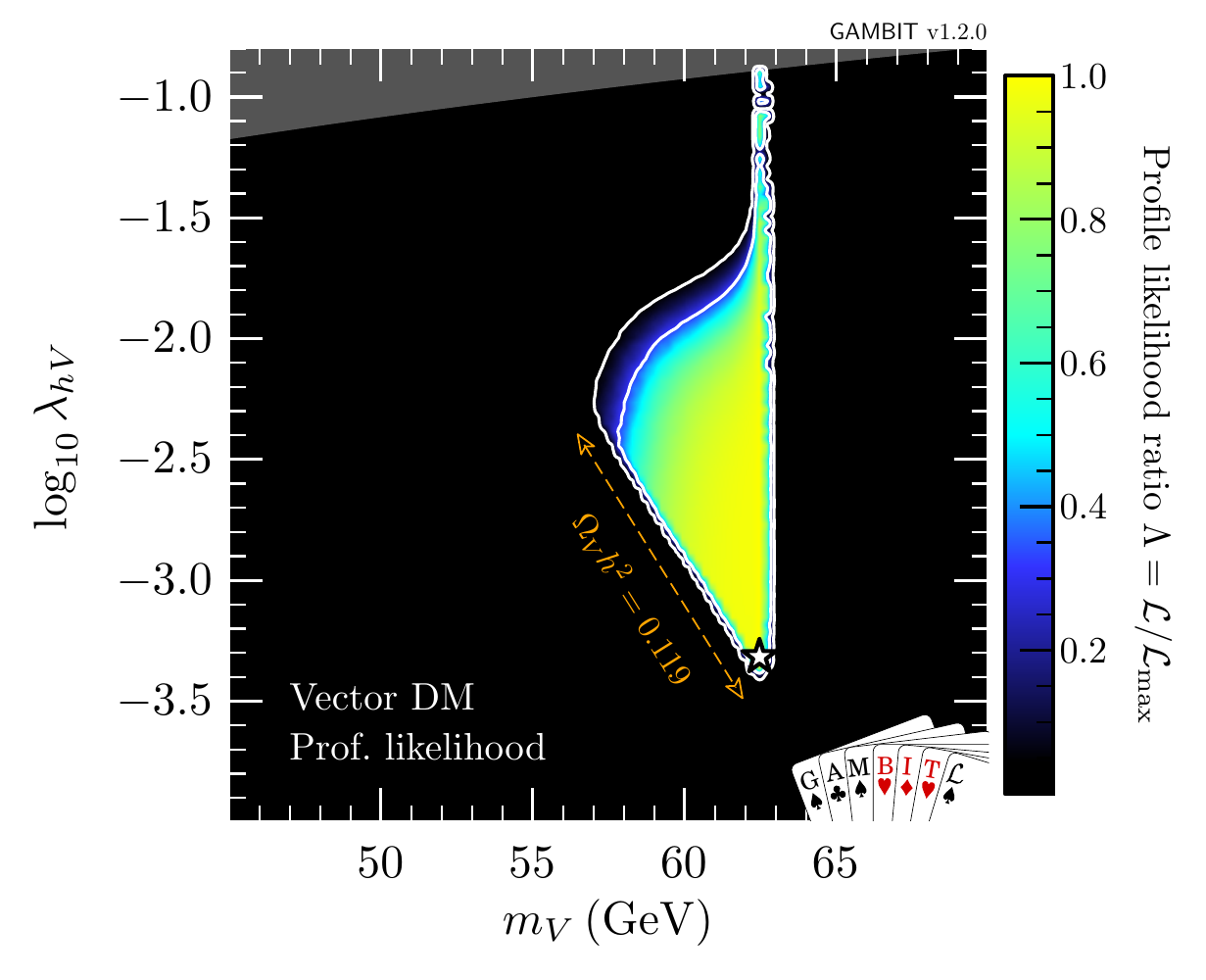} 
	\includegraphics[width=0.48\textwidth]{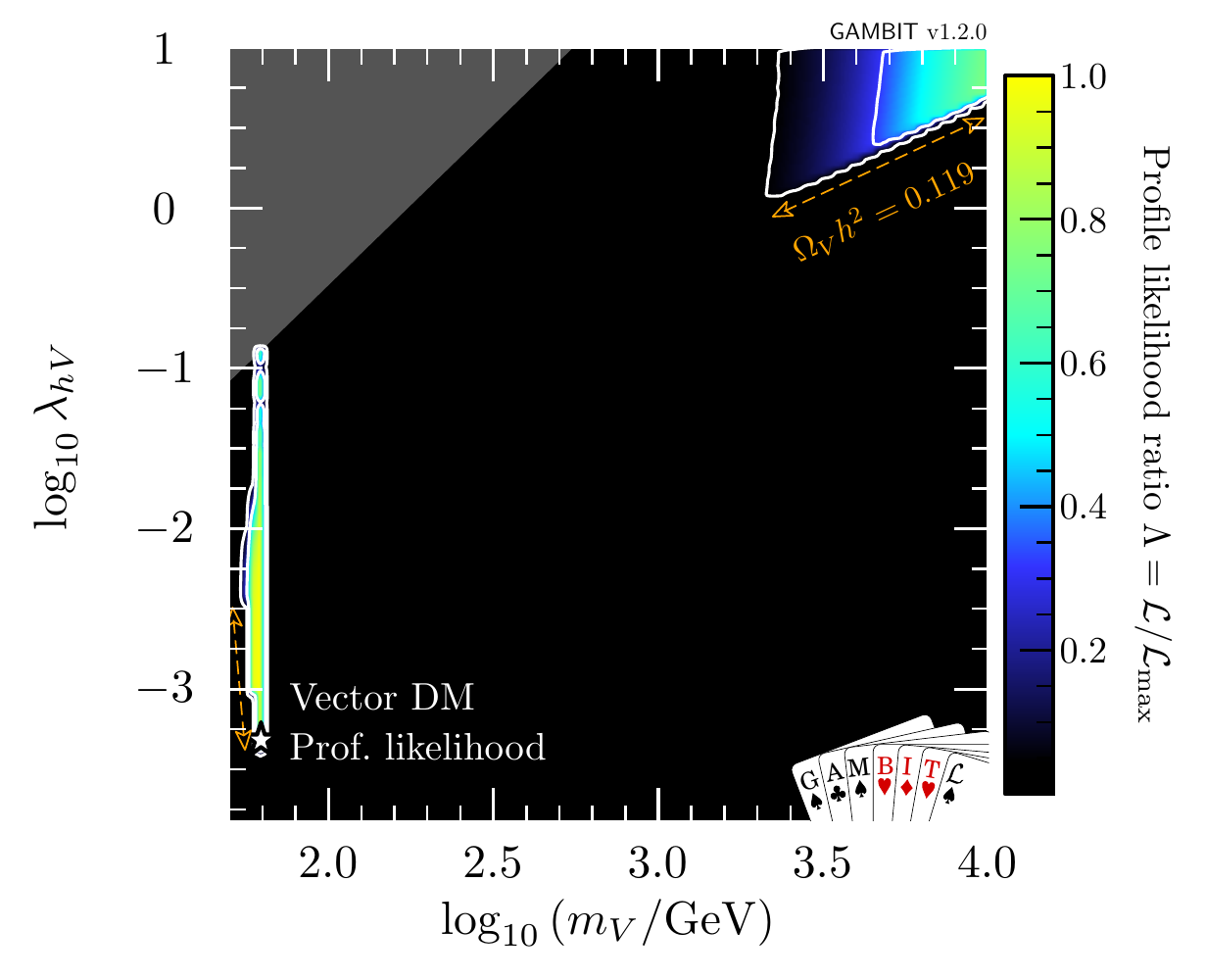}
	
	\caption{2D profile likelihoods in the $(m_V,\,\lambda_{hV})$ plane. The white curves show the $1\sigma$ and $2\sigma$ C.L. contours. The best-fit point is shown as a white star while the dark grey region is ruled out by perturbative unitarity constraint, $0 \leq \lambda_{hV} \leq 2 m_V^2/v_0^2$; see text for more details.}
	\label{fig:VDM_prof}
\end{figure}

In Fig.~\ref{fig:VDM_prof}, we show the profile likelihoods in the $(m_V,\,\lambda_{hV})$ plane. From these plots, we make the following observations.
\begin{itemize}
	\item In the low-mass range, the model parameter space is ruled out from below by the DM relic density, from upper-left by the Higgs invisible decay for $m_V < m_h/2$, and from right by indirect and direct detection limits. However, the resonance region around $m_V \sim m_h/2$ remains compatible as we do not require the vector DM to saturate the observed DM abundance. The `neck' region at $m_V \simeq m_h/2$ is excluded by the perturbative unitarity constraint. This is not true in the case of scalar singlet model \cite{Beniwal:2015sdl,Athron:2018ipf,Athron:2017kgt}.

	\item We find viable solutions for $m_V \gtrsim 10$\,TeV and $\lambda_{hV} \gtrsim 1$ which are constrained from below (left) by the DM relic density (direct detection limits). In contrast to the scalar singlet model where a second (allowed) region also appears at $m_V \simeq m_h$ \cite{Athron:2018ipf, Athron:2017kgt}, this is ruled out in the vector DM model by the perturbative unitarity constraint, see the dark grey region.
\end{itemize}

\begin{figure}[t]
	\centering
	
	\includegraphics[width=0.48\textwidth]{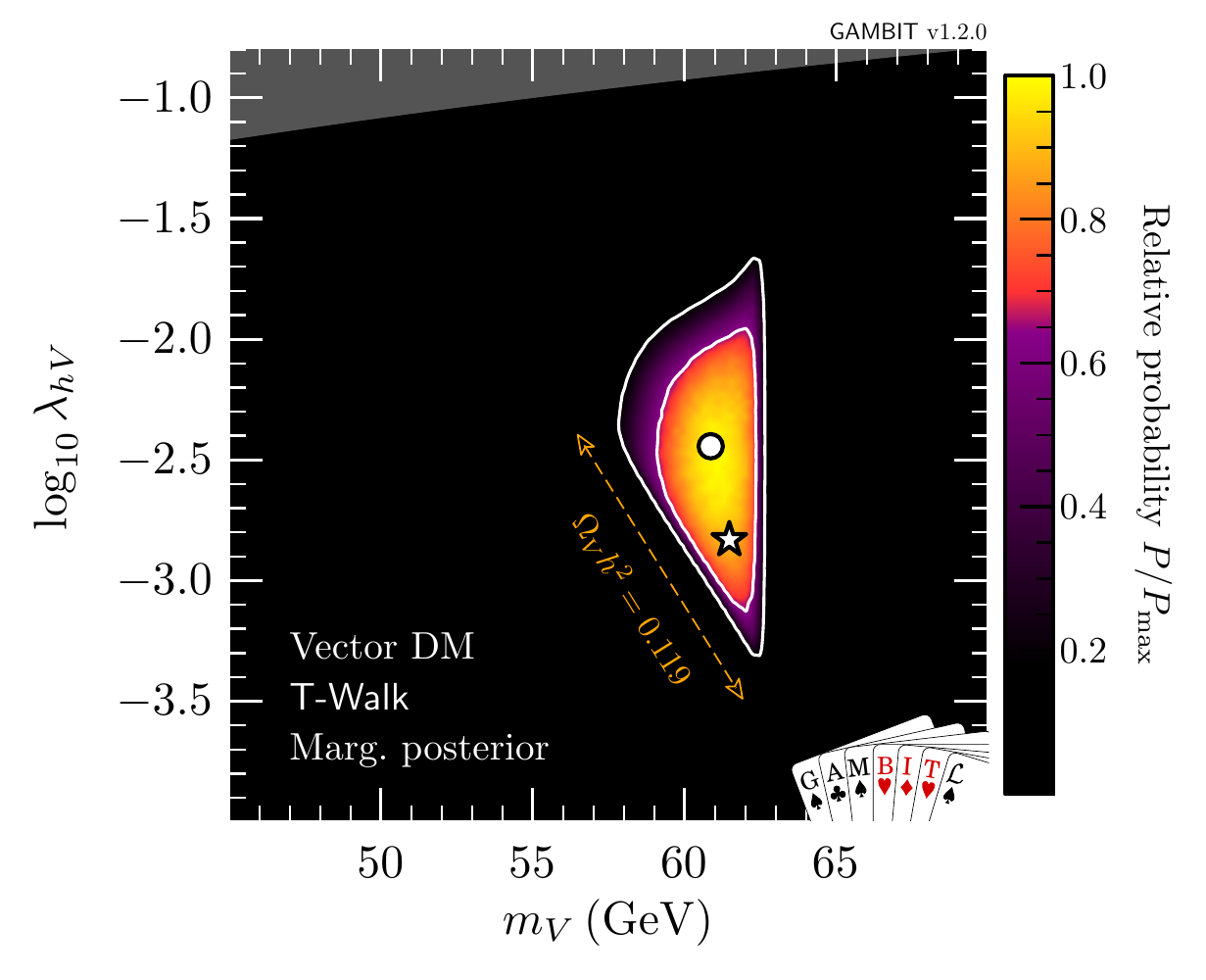} 
	\includegraphics[width=0.48\textwidth]{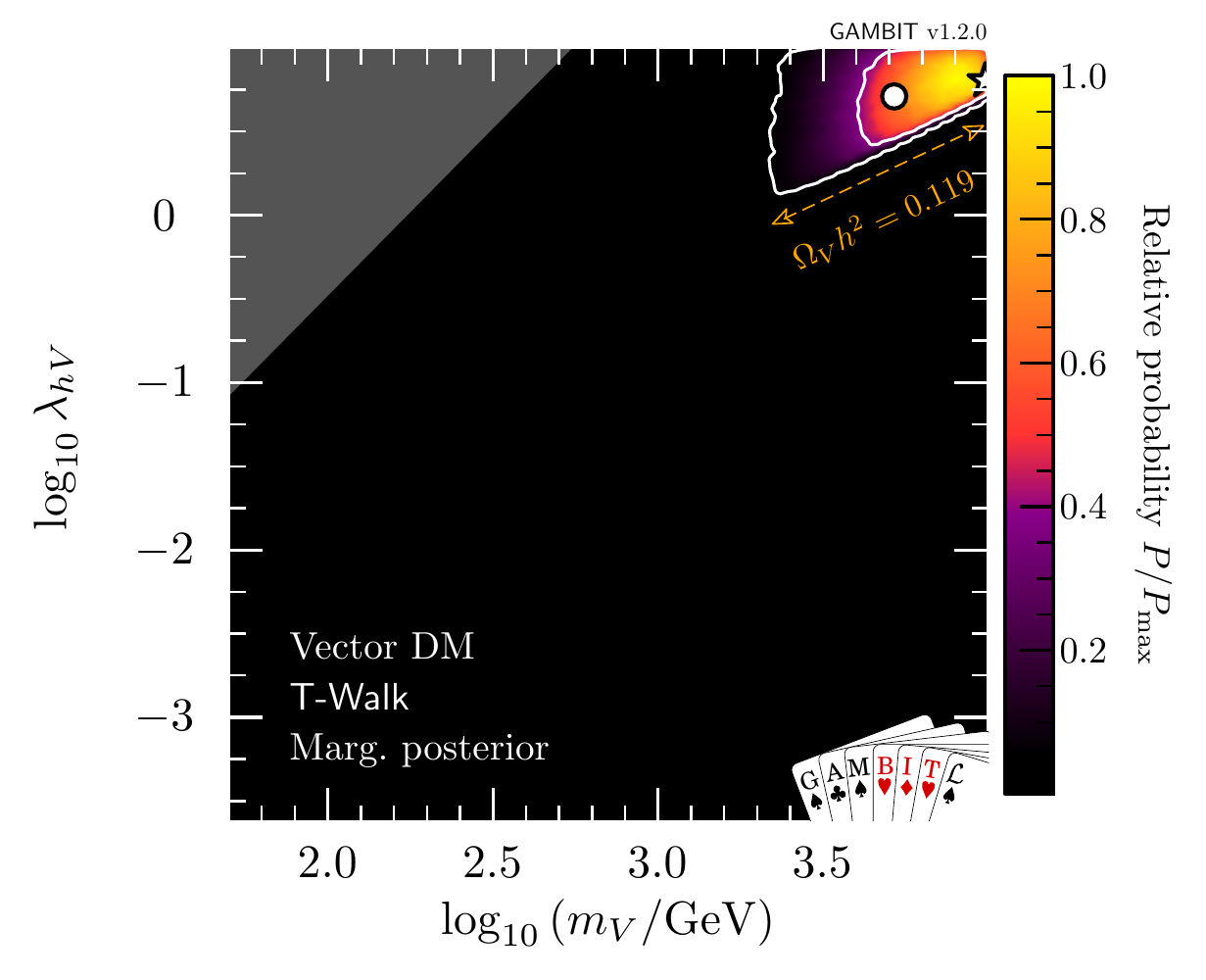}
	
	\caption{2D marginalised posterior distributions in the $(m_V,\,\lambda_{hV})$ plane.~The white curves show the $1\sigma$ and $2\sigma$ credible intervals. The posterior mean is shown as a white circle; the white star and dark grey region has the same meaning as in Fig.~\ref{fig:VDM_prof}.}
	\label{fig:VDM_marg}
\end{figure}

In Fig.~\ref{fig:VDM_marg}, we show the marginalised posterior distributions in the $(m_V,\,\lambda_{hV})$ plane.~Here the posterior mean is shown as a white circle.~With respect to Fig.~\ref{fig:VDM_prof}, the main differences are: (\emph{i}) the neck region in the low mass range is disfavoured after marginalising over nuisance parameters, particularly $m_h$. Thus, the allowed region is diluted due to volume effects; and (\emph{ii}) the resonance region in the high mass range is fine-tuned.~Although the best-fit point appears in the resonance region, the posterior mass is so small there that it falls outside the $2\sigma$ credible interval.

\subsection{Majorana fermion dark matter}
\begin{figure}[t]
	\centering
	\hspace{-5mm}
	
	\includegraphics[width=0.328\textwidth]{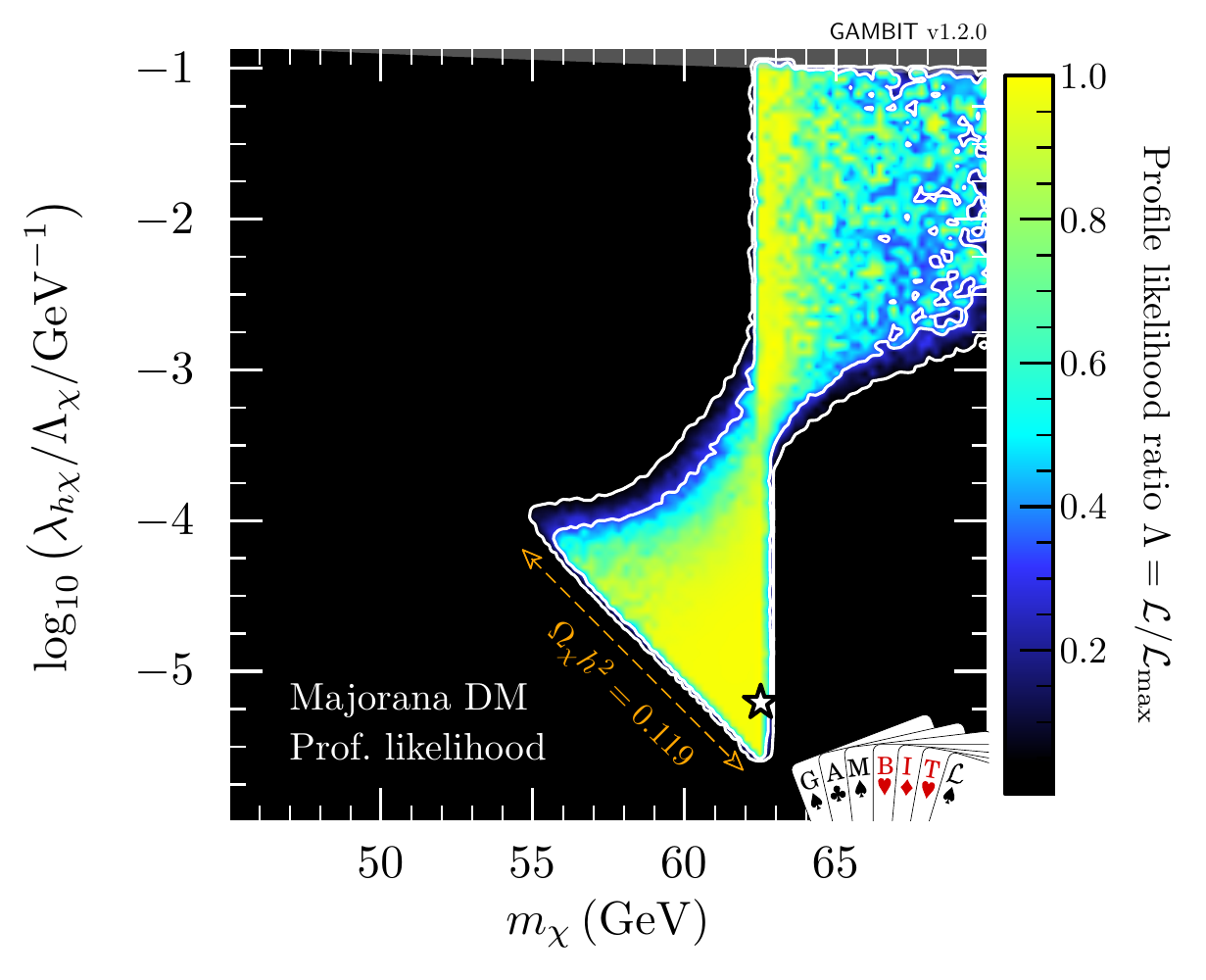}
	\includegraphics[width=0.328\textwidth]{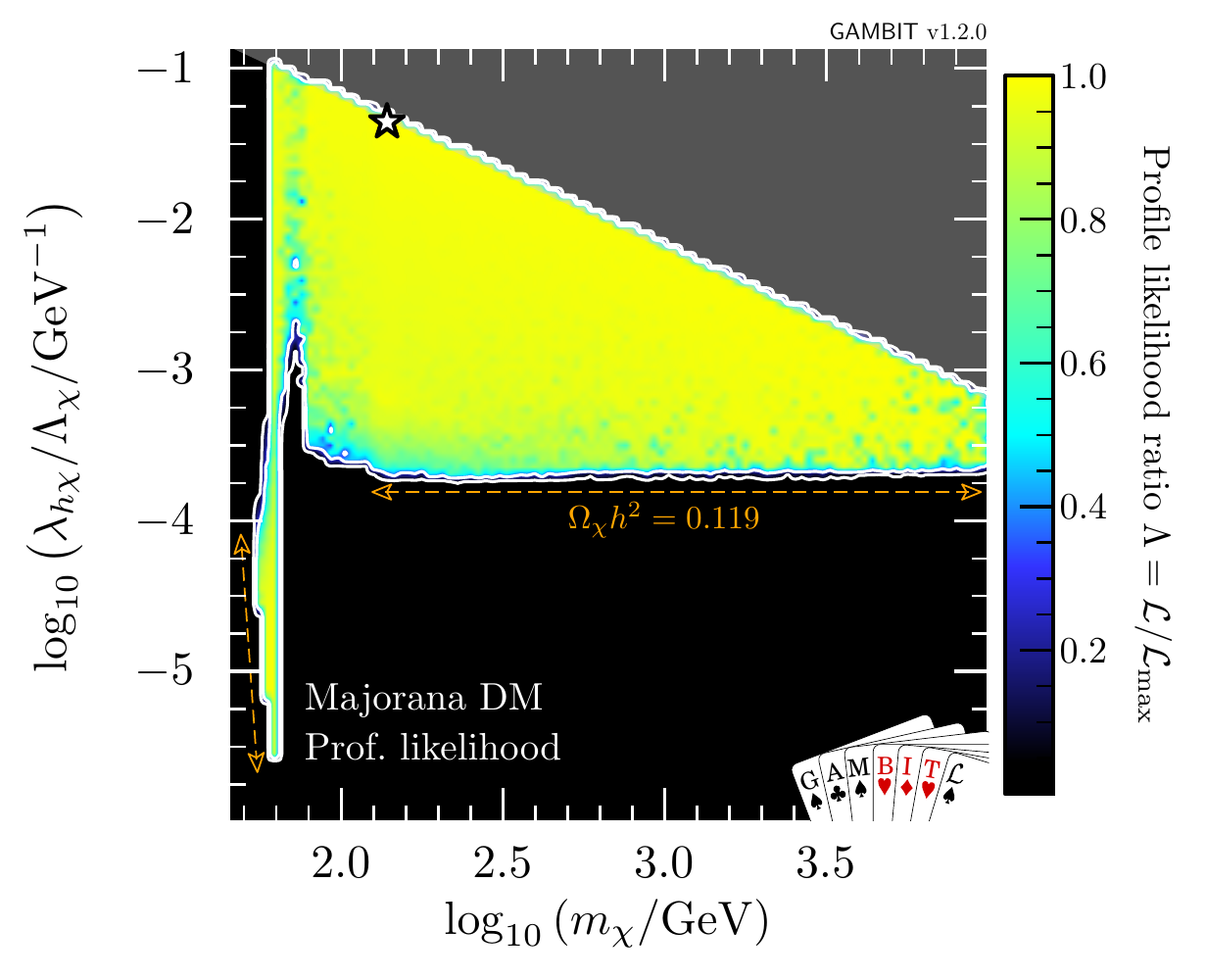}
	\includegraphics[width=0.328\textwidth]{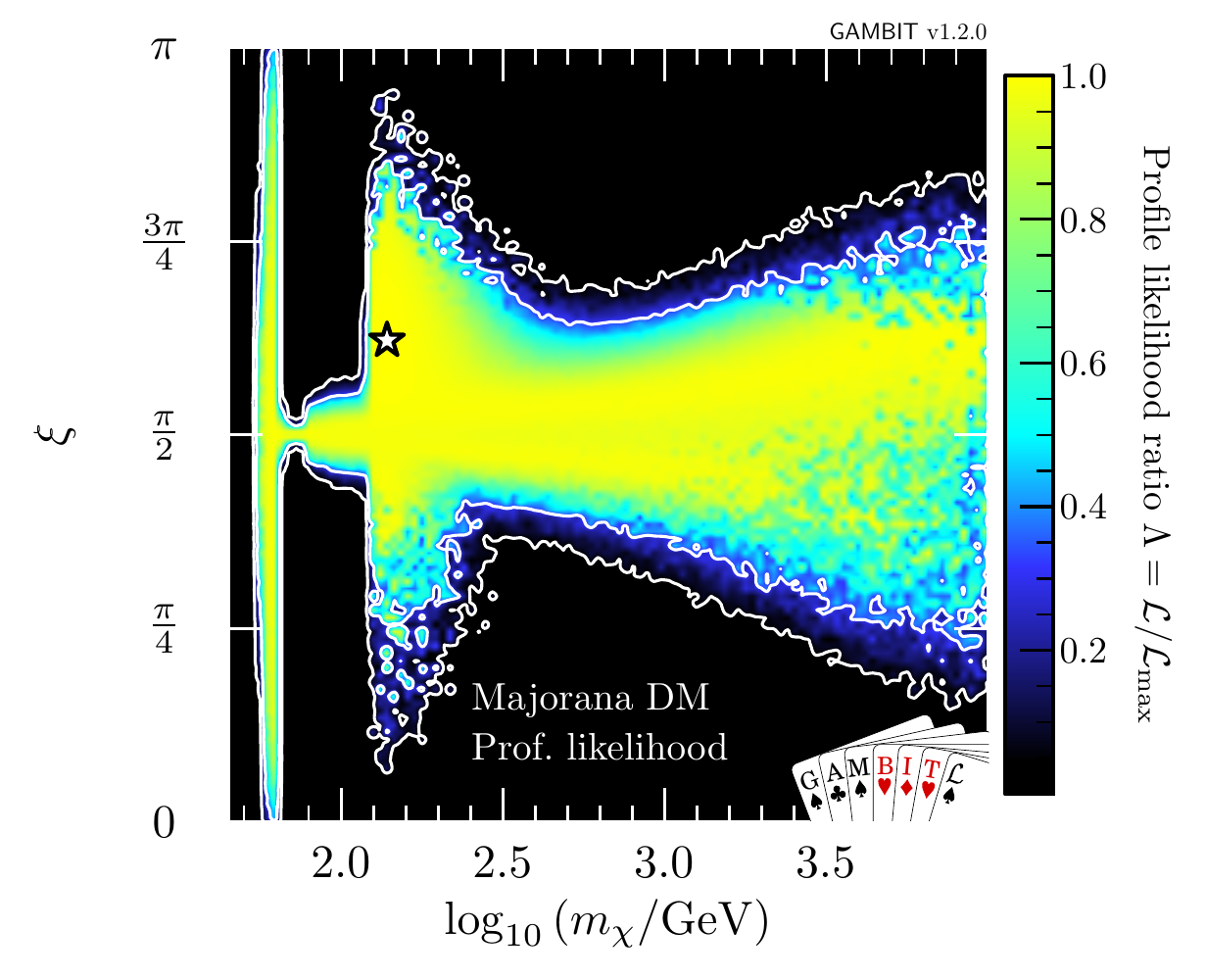}
	
	\caption{2D profile likelihoods in the planes of Majorana fermion model parameters.~The white curves show the $1\sigma$ and $2\sigma$ C.L. contours. The best-fit point is shown as a white star while the dark grey region is ruled out by the EFT validity constraint, $\lambda_{h\chi}/\Lambda_\chi < 4\pi/(2 m_\chi)$; see text for more details.}
	\label{fig:MDM_prof}	
	
	\hspace{-5mm}
	
	\includegraphics[width=0.328\textwidth]{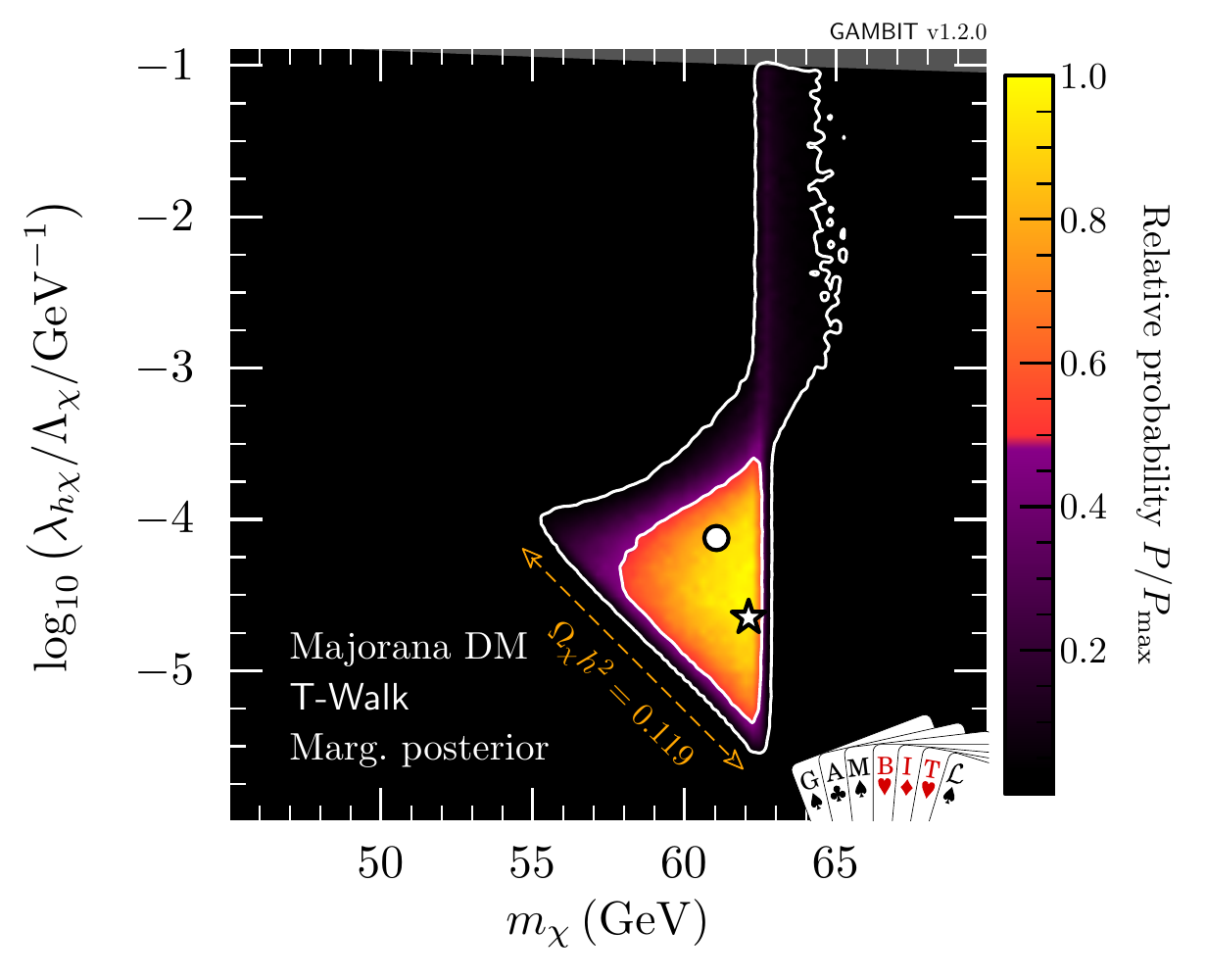}
	\includegraphics[width=0.328\textwidth]{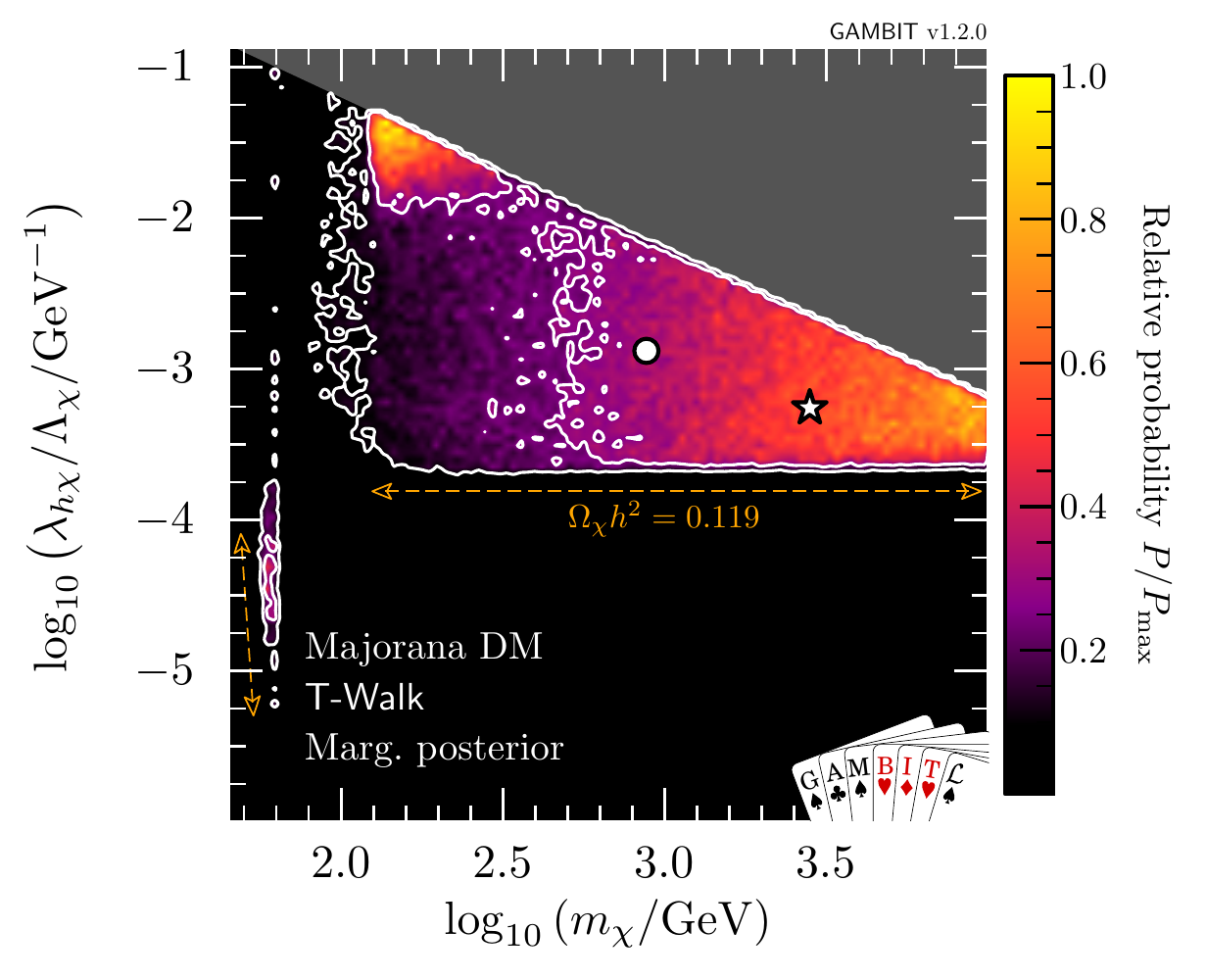}
	\includegraphics[width=0.328\textwidth]{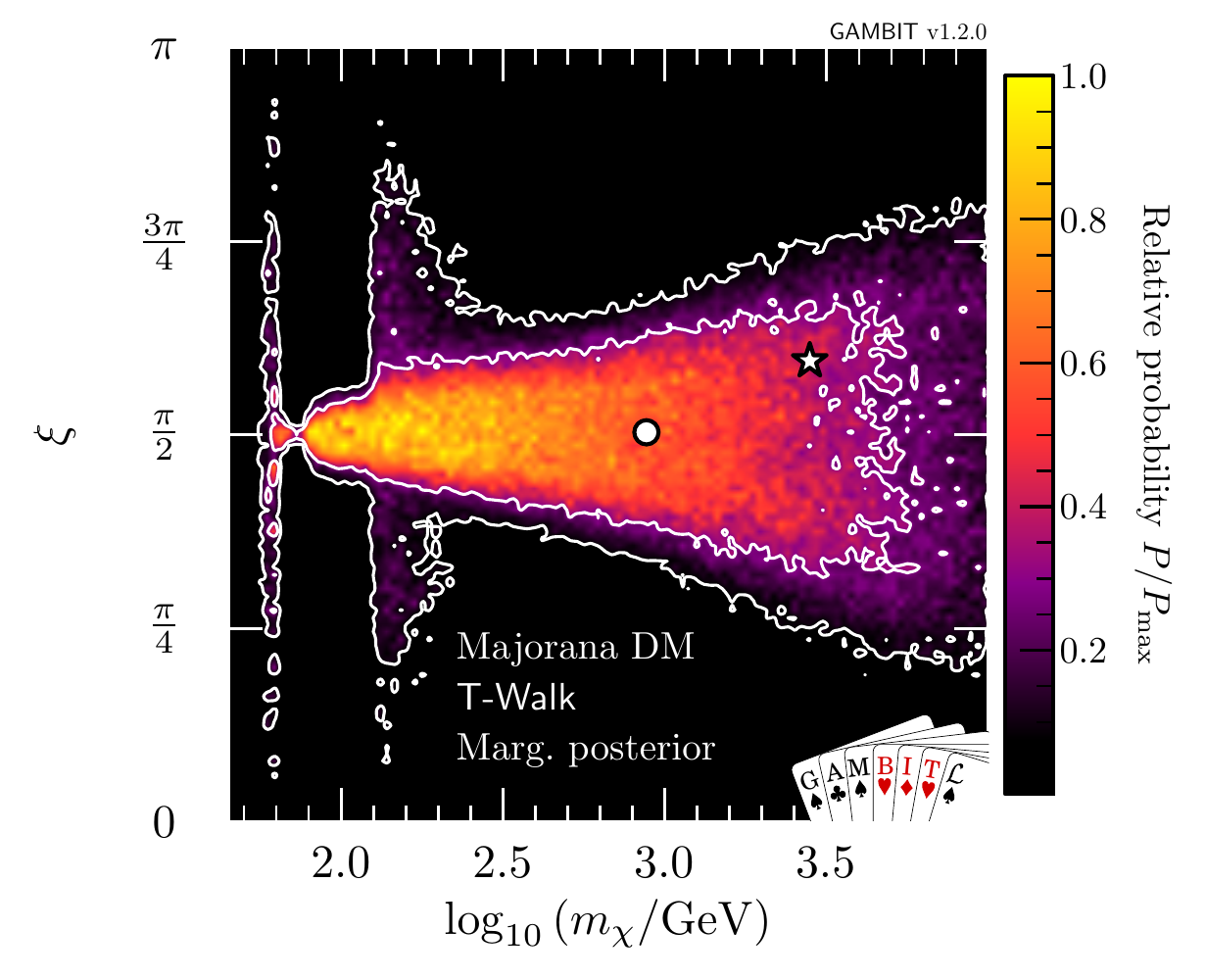}

	\caption{2D marginalised posteriors in the planes of Majorana fermion model parameters.~The white curves show the $1\sigma$ and $2\sigma$ credible intervals.~The posterior mean is shown as a white circle; the white star and dark grey region has the same meaning as in Fig.~\ref{fig:MDM_prof}.}
	\label{fig:MDM_marg}
\end{figure}

The profile likelihoods for the Majorana fermion DM model are shown in Fig.~\ref{fig:MDM_prof}. From these plots, we make the following observations.
\begin{itemize}
	\item For the low mass range in the $(m_\chi,\,\lambda_{h\chi}/\Lambda_\chi)$ plane, the allowed parameter space is similar to the vector DM model.~However, in contrast, the low and high mass solutions are now connected. This is entirely due to the parameter $\xi$, which is profiled over in this plane. As a pure-pseudoscalar coupling, $\xi = \pi/2$, suppresses the direct detection limits, see Eq.~\eqref{eqn:suppXsection}, the two regions are connected -- this is also evident in the $(m_\chi,\,\xi)$ plane.

	\item The high mass region in the $(m_\chi,\,\xi)$ plane prefers $\xi \sim \pi/2$, however, larger variations in $\xi$ are allowed for large $m_\chi$ -- here the direct detection limits become less constraining. In the $(m_\chi,\,\lambda_{h\chi}/\Lambda_\chi)$ plane, fermion DM masses between 100 GeV and 10 TeV are allowed with a degenerate likelihood once $\xi$ is profiled out, otherwise the parameter space is ruled out from below (above) by the DM relic density (EFT validity constraint, see section~\ref{sec:theory}).
\end{itemize}

In Fig.~\ref{fig:MDM_marg}, we show the marginalised posterior distributions in the planes of Majorana fermion model parameters.~The posterior mean (best-fit) is shown as a white circle (star).~Similar to the profile likelihood plots in Fig.~\ref{fig:MDM_prof}, the free mixing parameter $\xi$ leads to a substantially larger allowed parameter space.~In the low mass range, the neck region is similar to the one in scalar and vector DM models, however it lies within the $2\sigma$ credible region due to a suppression of direct detection limits when $\xi \sim \pi/2$ -- this softens the penalty from marginalisation over the nuisance parameters. When the posterior is computed over the high mass range, the resonance region is somewhat disfavoured, but a larger parameter space opens up for $m_\chi \gtrsim m_h$. This is also evident in the $(m_\chi,\,\xi)$ plane.

\section{Summary}
We have performed a global study of the effective vector and Majorana fermion Higgs portal models using the \textsf{GAMBIT} software.~Using the latest results from \emph{Planck} measured DM relic density, Higgs invisible decay, and indirect and direct detection experiments, we systematically explored the parameter space of these models using state-of-the-art scanning tools in both frequentist and Bayesian frameworks. We also ensured that the vector (Majorana fermion) DM models satisfy the perturbative unitarity (EFT validity) bounds.~In addition, we included a handful of SM, nuclear physics, DM halo and velocity distribution as nuisance parameters, and profiled (marginalised) over them to produce $1\sigma$ and $2\sigma$ C.L. (credible intervals) in various planes of the model parameters.

Our results showed that the resonance region is consistent with all experimental constraints. On the other hand, the high mass range in our models is constrained by a combination of DM relic density, direct detection and theoretical constraints.~In this regards, the Majorana fermion model with a CP-violating coupling $(\xi \sim \pi/2)$ is most interesting -- this choice leads to a momentum-suppression of the DM-nucleon cross-section, and thus suppresses the direct detection limits. Consequently, a large portion of the parameter space opens up that can potentially be probed using future indirect searches.

\vspace{1.5mm}
\noindent \textbf{Note}: Here we have only presented a subset of our full global fit results; for more details, including the results for the Dirac fermion model, Bayesian model comparison \emph{etc}., see Ref.~\cite{Athron:2018hpc}.~All of our samples and input files are publicly available via \href{https://www.zenodo.org/communities/gambit-official/}{\textsf{Zenodo}}.

\acknowledgments{AB gratefully acknowledges all members of the GAMBIT Collaboration for their involvement in this project, and the EPS-HEP 2019 organisers/convenors for the opportunity to present this work.~AB is supported by F.N.R.S. through the F.6001.19 convention.}

\twocolumn
{
	\bibliographystyle{JHEP}
	\bibliography{EPSHEP2019_Ankit.bib}	
}

\end{document}